\documentclass[useAMS,usenatbib]{elsart}
\usepackage{times}
\usepackage{epsfig}
\usepackage{graphicx}
\usepackage[usenames,dvips]{color}

\newcommand{\commentout}[1]{}
\newcommand{\eps}{\varepsilon}
\newcommand{\src}[1]{}

\begin{document}

\begin{frontmatter}



\title{Gravitational Softening and Adaptive Mass Resolution}


\author{Alexander Shirokov}

\address{CITA, University of Toronto, 60 St. George Street, Toronto, Ontario, M5S 3H8}

\begin{abstract}
  Pairwise forces between particles in cosmological N-body simulations are
  generally softened to avoid hard collisions.  Physically, this softening
  corresponds to treating the particles as diffuse clouds rather than point
  masses. For particles of unequal mass (and hence unequal softening length),
  computing the softened force involves a nontrivial double integral over the
  volumes of the two particles.  We show that Plummer force softening is
  consistent with this interpretation of softening while spline softening is
  not.  We provide closed-form expressions and numerical implementation for
  pairwise gravitational force laws for pairs of particles of general
  softening scales $\varepsilon_1$ and $\varepsilon_2$ assuming the commonly
  used cloud profiles: NGP, CIC, TSC, and PQS. Similarly, we generalize
  Plummer force law into pairs of particles of general softenings.  We relate
  our expressions to the gaussian, Plummer and spline force softenings known
  from literature.  Our expressions allow possible inclusions of pointlike particles
  such as stars or supermassive black holes.
\end{abstract}


\end{frontmatter}


\section{Introduction}\label{sec_intro}

Gravitational softening is a building block in the foundation of any modern
cosmological N-body simulation; at the same time it has traditionally been
given relatively little coverage in literature. In this paper we give this
problem a share of our dedicated attention.

Any practical numerical simulation will have finite dynamic range, limiting
the resolution with which the system of interest may be studied.  Oftentimes,
the resolution requirements for simulations are not completely uniform across
the simulation volume: there may be special regions of interest, which we
would like to study at high resolution, whereas the remainder of the
simulation volume need not be simulated in great detail.  For example, we
might be interested in simulating individual dark matter halos embedded within
realistic, time-varying cosmological environments (e.g.  \cite{Diemand05},
\cite{Diemand07}).

Given finite computational resources, an efficient strategy is to employ
adaptive resolution, i.e.\ to use high resolution in the regions of interest
and low resolution elsewhere \cite{Bertschinger01}.  Two types of resolution that arise
in N-body simulations are mass resolution and force resolution.  The mass
resolution is simply limited by the finite number of particles; increasing the
particle number within some region increases its mass resolution.  Force
resolution is limited by softening of pairwise forces between particles.  For
simulations of collisionless systems, pairwise forces must be softened at
short distances to avoid artificial collisionality due to finite particle
number (e.g. \cite{white78}, \cite{power03}, \cite{lukic07} and references therein).  For
example, Plummer softening corresponds to calculating the force between two
particles separated by $\vec{r}$ as
\begin{equation}\label{eq_plummerforce}
\vec{F}(\vec{r})=-\frac{m_1 m_2}{(|r|^2 + \eps_P^2)^{3/2}}\vec{r},
\label{plummer}
\end{equation}
where $\eps_P$ here is the Plummer softening length.  
Other softening
laws are commonly used in the literature, with their own corresponding
softening lengths.  
Increasing the force resolution therefore means decreasing the force
softening length.  This is allowed within regions of high mass
resolution where the particle number is enhanced; elsewhere, the force
softening length must remain large to avoid collisionality.  One
common choice is to scale the softening length as the cube root of the
mass, $\eps\propto m^{1/3}$, which holds fixed the maximal density of
all particles.

In Eqn.~(\ref{plummer}), one subtlety that arises when computing the
pairwise force between particles of different mass is that the
appropriate choice of softening length becomes uncertain: should the
force be softened by the larger $\eps$, the smaller, or some average
of the two?   The chief requirements are that the
pairwise forces be symmetric in order to conserve momentum (i.e.\ the
same $\eps$ is used for $F_{12}$ as for $F_{21}$); and that the
softening length is not too small, in order to avoid hard collisions.
For example, taking $\eps={\rm max}(\eps_1,\eps_2)$ satisfies both of
these conditions.  This choice is clearly not optimal, however, since
the effective force resolution is degraded. 

In addition to basic conservation laws, the consistency of computed forces
with gravity is also important. Consider a set of point masses connected by
massless springs for example: such system clearly satisfies both energy and
momentum conservation laws but its evolution is inconsistent with gravity. This illustrates
that in setting up the effective softening one should also ensure the
consistency of computed forces with gravity.

The uncertainty in the appropriate softening method may be resolved 
by resorting to a well-known physical interpretation of force
softening.  The appropriate physical model is that simulation particles
represent not point masses, but rather spatially extended clouds of
finite density.  This picture naturally leads to
finite maximal pairwise forces : if $\rho\rightarrow\rho_{\rm max}$
towards the center of a cloud, then the force exerted on a test
particle will reach a maximum and eventually vanish as the test
particle approaches the interior and then the center of the cloud.  
Of course, to be self-consistent within this physical picture, we
should not compute pairwise forces between particles treating one
particle as an extended cloud and the other as a point particle, but
rather treating both particles as extended.  The pairwise interaction
potential then becomes 
\begin{equation}\label{correct}
U_{12} = -\int
\frac{\rho_1(\vec{x_1})\rho_2(\vec{x_2})}{|\vec{x_1}-\vec{x_2}|}\;
d^3x_1 d^3x_2,
\end{equation}
and the force is obtained by differentiation of the potential with respect to
separation between the centroids of clouds~1 and 2.

Equation~(\ref{correct}), while clearly the correct choice of force softening,
may appear daunting to evaluate.  Pairwise force calculation is often the
limiting step in N-body simulations, and so an expression for the pairwise
force that requires few floating point operations in its evaluation is an
absolute requirement.  The 6-dimensional integral in Eqn.~(\ref{correct})
would hardly appear promising in this regard, and so a simpler choice of
softening, although less physically self-consistent, may seem more appealing.
However, we show in this paper that (perhaps surprisingly), the potential in
Eqn.~(\ref{correct}) is in fact analytic for commonly used softening kernels.
For the popular ``spline'' density kernel \cite{Monaghan85}, the potential in
Eqn.~(\ref{correct}) may be expressed with rational functions, while for
Plummer force kernel the integral can be done numerically. This allows the
efficient calculation of self-consistent forces between particles of unequal
mass using our proposed W-shape and the extended Plummer force laws.  Our
force profiles can also be effectively used for particle splitting methods
\cite{mes05} applied to pure gravity, or simulation of dark matter including
pointlike objects such as super-massive black holes or stars.

Section~\ref{sec:review} gives general expressions for pairwise gravitational
potential and force given the specified density shape for particles in the
pair. A numeric integration is necessary for Plummer softening discussed in
section~\ref{sec_plummer}. Closed form analytic solution for standard $W_n$
shapes for~$n=1,2,3,4$ is presented in section~\ref{sec:softening}. As
$n\rightarrow \infty$, this solution asymptotically approaches gaussian
softening discussed in section~\ref{sec_gaussian}.  In section~\ref{sec_relat}
we compare all the discussed softening methods and relate them to each other. We conclude in
section~\ref{sec:concl}.

\section{Review of Pairwise Forces}\label{sec:review}

Consider a simulation particle at position $\vec{x}_0$, of mass $m$
and density profile $\rho(\vec{x})=m\;W(\vec{x}-\vec{x}_0)$, where the
cloud shape $W$ has unit normalization:
\begin{equation}
\int W(\vec{x})d^3\vec{x}=1.
\end{equation}
We will consider only spherically symmetric shapes,
$W(\vec{r}\,)=W(r=|\vec{x}-\vec{x}_0|)$ to avoid generation of spin angular momenta
from tidal torques.  Then the Fourier transform of the shape similarly
depends only on the magnitude $k$ of the wavevector~$\vec{k}$, 
\begin{equation}\label{eq_wreal}
  \begin{array}{ccc}
    W(\vec{k})&=&W(k)=\int W(r) e^{-i\vec{k}\cdot\vec{r}}d^3\vec{r}\\[5pt]
    &=& 4\pi\int W(r) \frac{\sin(kr)}{kr} r^2 dr\;,
  \end{array}
\end{equation}
and is a real function.  

Given the particle's density profile, its gravitational potential is
implicitly defined by
\begin{equation}
  \nabla^2\varphi(\vec{x}) = 4\pi \rho(\vec{x}),
\end{equation}
where we set Newton's gravitational constant $G_N=1$ for simplicity.
We can write an explicit expression for the potential using the Greens
function for the Laplacian operator:
\begin{equation}
  \varphi(\vec{x}) = \int  G_\varphi(\vec{x},\vec{y}\,)\,
  \rho(\vec{y}\,)\; d^3 \vec{y}\;,
\end{equation}
where the Greens function satisfies 
\begin{equation}\label{eq_gphi}
  \nabla^2 G_\varphi(\vec{x},\vec{y}) = 4\pi \delta^{(3)}(\vec{x}-\vec{y}\,)\;.
\end{equation}
Then, consistently with Eqn.~(\ref{correct}), we can write down the pairwise
interaction energy between particles $i$ and~$j$ as
\begin{equation}\label{eqn:poten}
  \begin{array}{c}
\displaystyle
    U_{ij}(\vec{r}_{ij}) = U_{ij}(r_{ij}) = \int \rho_{i}(\vec{x})\,\varphi_j(\vec{x}) \;d^3x 
\\
\displaystyle
    = m_i m_j \int W_i^*(k) W_j(k) G_\varphi(k)\, e^{i\vec{k}\vec{r}_{ij}} \frac{d^3 \vec{k}}{(2\pi)^3} 
\\
\displaystyle
    = -\frac{2 m_i m_j}{\pi} \int_0^\infty W_i^*(k) W_j(k) j_0(kr_{ij})dk\,, \nonumber
  \end{array}
\end{equation}
where we have used $G_\varphi(k)=-4\pi/k^2$, $j_0(x)=\sin(x)/x$, and 
$r_{ij}=|\vec{r}_{ij}|$, $\vec{r}_{ij}=\vec{x}_i-\vec{x}_j$.  

Differentiating with respect to the separation vector and flipping the sign
gives pairwise force
\begin{eqnarray}\label{eqn_force}
  \vec{F}_{ij}(\vec{k}) = -i\vec{k}\, W_i^*(k) W_j(k) G_\varphi(k)\;,
\end{eqnarray}
 applied on particle~$i$ by particle~$j$.

As an example, we can consider point particles, with
$W(\vec{r})=\delta^{(3)}(\vec{r})$, or equivalently $W(k)=1$.  Then
Eqn.~(\ref{eqn:poten}) gives $U_{ij}=-m_im_j/r_{ij}$ or force $\vec{F}_{ij} =
-m_im_j\vec{r}_{ij}/r_{ij}^3$.  More
interesting applications will arise when we consider some other commonly used
cloud profiles, discussed in the following sections.

\section{Plummer Softening}\label{sec_plummer}

\begin{figure}
\includegraphics[scale=0.6]{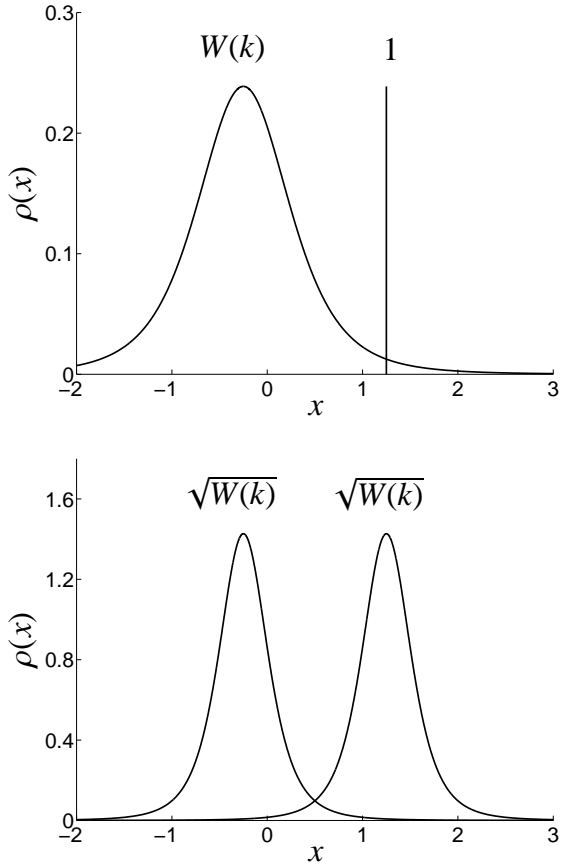}
\caption{\label{fig_plummerden} Force softening such as Plummer can be viewed
  as force between a cloud of shape $W(k)\ge 0$ and a point mass (shown in
  physical space on {\it top plate}), or alternatively as force between two
  identical cloud shapes $\sqrt{W(k)}$ {\it (bottom plate)}. We plot Plummer
  shapes with $\sqrt{W(k)} \equiv W_P(\eps_P,k)$ in this example, for $\eps_P=1$.
}
\end{figure}

Plummer softening Eqn.~(\ref{plummer}) leads to fourier component
$\vec{F}_P(\vec{k}) = m_i m_j 4\pi i \vec{k} \eps_P K_1(k \eps_P)/k$, where
$K_\nu(x)$ is the modified Bessel function (using Eqns.~3.771.2 and 3.771.5
of \cite{Gradshteyn94}).  Comparing this with Eqn.~(\ref{eqn_force}) we find
$W_i^*(k)W_j(k) = k\eps_P K_1(k\eps_P)$.  

This implies that Plummer force can be viewed as force between a cloud shape
whose fourier component is $W_i(k)=k\eps_P K_1(k\eps_P)$ and a point mass~$W_j(k)=1$.
Alternatively, Plummer force can be viewed as force between two identical
Plummer cloud shapes
\begin{equation}\label{plummer:w}
  W_P(\eps_P,k) = \sqrt{k\eps_P K_1(k\eps_P)}\;.
\end{equation}
Figure~\ref{fig_plummerden} illustrates
these alternatives in physical space perspective.

Plummer force law can be consistently extended into pairs of particles on
unequal softening by using $W_P(\eps_1,k)$ and $W_P(\eps_2,k)$ in
Eqn.~(\ref{eqn:poten}) for computing potentials and forces for all pairs of
particles. The integral leads to the Plummer force Eqn.(\ref{plummer}) for
pair of identical particles $\eps_1 = \eps_2 = \eps_P$. Numerical evaluation
is required however for pairs of unequal softenings. Our implementation in
Appendix~\ref{sec_numeric} includes potentials and forces for these Plummer
shapes. 

As an alternative method of extending Plummer law into pairs of unequal
softening, one would consider plugging a symmetric combination of $\eps_1$
and~$\eps_2$ into Eqn.(\ref{plummer}). However, as we show in
sections~\ref{sec_gaussian} and~\ref{sec_relat} such an approach leads to
inconsistency of the resulting forces with gravity for pairs of unequal softenings.

\section{Softening with $W_n$ Clouds}\label{sec:softening}

The assignment of particle mass to a regular grid is an important step in
particle-mesh codes \cite{he88}, and is achieved using one of several
possible shapes:\ Nearest Grid Point (NGP), Cloud in Cell (CIC),
Triangular-Shaped Cloud (TSC). In this section we discuss how these shapes are
used to define our proposed W-shape softening and spline force softening
previously used in literature.

\subsection{Hockney-Eastwood Cloud Shapes}

We can write the Hockney-Eastwood shapes in one dimension as
\begin{equation}\label{eq_WaP}
  w_n(s) = 
  \frac{1}{\pi}
  \int\limits_0^\infty
  \cos\,(k_s s)\left(\frac{\sin k_s/2}{k_s/2}\right)^n\;dk_s\;,
\end{equation}
The $w_n(s=nx/b)$ clouds are characterized by two quantities: the scale length
$b$, and the index $n$ which controls the smoothness of the function.  The
function $w_n(s)$ has $n-1$ continuous derivatives and disappears at $x > b/2$.

These Hockney-Eastwood cloud shapes are defined in one dimension (see
Table~\ref{he-clouds}), however we can generalize them by replacing
their argument $s$, a linear coordinate, with spherical radius $r$.
Let us define
\begin{equation}
  W_n(b,r)=\frac{6n^2}{\pi b^3} w_n(n\,r/b)
\label{W_defn}
\end{equation}
where the prefactor is inserted to ensure that $W_n$ is properly normalized.
Because these cloud shapes have compact support, the force law they generate
on a point test particle is exactly Newtonian for $r>b/2$.

Note that $n=4$ corresponds to the so-called `spline' softening used, for
example, in smoothed particle hydrodynamics \cite{Monaghan85} and pure gravity.
Using $b=2h$ as an exercise identically yields the density kernel in Eqn.(4) of
\cite{Springel05}; using $b=4h$ yields Eqn.(11) of \cite{Price07}.

\begin{table}
\begin{displaymath}
  \begin{array}{@{\extracolsep{2mm}}|c|c@{\quad}|r@{\quad}|}
\hline
&&\\[-20pt]
   \mbox{\hspace{1mm}Scheme\hspace{1mm}}& n & {w}_n(s), s>0
\\[5pt]
\hline
&&\\[-15pt]
   \mbox{NGP}   & 1 & 1-H_{1/2}(s)
\\[-2pt]
&&\\[-15pt]
   \mbox{CIC}   & 2 & (1-s)-(1-s)\,H_1(s)
\\[-2pt]
&&\\[-15pt]
   \mbox{TSC}   & 3 & (\frac{3}{4}-s^2) + \frac{3(2s-1)^2}{8} H_{1/2}(s) 
\\[1pt]
                &   &  - \frac{(2s-3)^2}{8} H_{3/2}(s)
\\[1pt]
&&\\[-15pt]
   \mbox{PQS}   & 4 & \left(\frac{2}{3} -s^2 + \frac{s^3}{2}\right)-\frac{2(s-1)^3}{3}H_1(s)
\\
                &   & +\frac{(s-2)^3}{6}H_2(s)
\\[3pt]
\hline
  \end{array}
\end{displaymath}
\caption{Hockney and Eastwood \cite{he88} cloud shapes in one dimension.  Here, $H_y(x)
  \equiv H_0(x-y)$, where  $H_0$ is the Heaviside function
defined as $H_0(x) = 0$ for $x < 0$ and $H_0(x)=1$ for $x > 0$.
\label{he-clouds}}
\end{table}

To compute the interaction potential, we require an expression for the
Fourier transform of $W_n(b,r)$ to insert into Eqn.~(\ref{eqn:poten}).
Using Eqns.~(\ref{eq_wreal}), (\ref{W_defn}) and (\ref{eq_WaP}), we find 
\begin{equation}\label{eq_ftwk}
W_n(b,k)={\cal W}_n(kb/2n)
\end{equation}
where
\begin{equation}\label{eq_ftwkfull}
\mathcal{W}_n(x)\equiv\frac{3 j_1(x)}{x}
\left(\frac{\sin x}{x}\right)^{n-1}\;,
\label{ft_wn}
\end{equation}
and $j_1(x) = (\sin{x}-x\cos{x})/x^2$ is the spherical Bessel function.
As an exercise, using $n=2$ and $b\equiv a$ one identically recovers Eqn.(A.16) of \cite{fb94}.

We have only verified this expression Eqn.~(\ref{ft_wn}) for $1\leq n\leq 4$, however
it may serve as our definition of the cloud
shape\footnote{Alternatively, one could define
  $\mathcal{W}_n(x)\equiv\left(\frac{3 j_1(x)}{x}\right)^n$ which is
  consistent with taking a 3-dimensional convolution.} for an arbitrary $n$.
Given this expression for the smoothing kernel, the interaction potential for
particles with smoothing scales $b_1$ and~$b_2$  becomes
\begin{equation}\label{eq_force_b1b2}
  \begin{array}{c}
    \displaystyle
    u_n(b_1,b_2,r)  = \frac{U_{ij}}{m_im_j} = \\[3mm]
    \displaystyle
    -\frac{2}{\pi} \int_0^\infty j_0(kr) \;
    \mathcal{W}_n\left(\frac{kb_1}{2n}\right)
    \mathcal{W}_n\left(\frac{kb_2}{2n}\right) 
    dk\;.
  \end{array}
\end{equation}

These integrals may be evaluated in closed form; the resulting expressions are
lengthy and given in the appendix.  For finite $n$, the force profile is
Newtonian ($f\propto r^{-2}$) outside $r>(b_1+b_2)/2$, and vanishes linearly
($f\propto r$) at the limit of zero separation.

Interaction potential has a simpler form in the case of a pair consisting of
two identical particles of smoothing scales~$b$
\begin{equation}\label{eq_force_b:2}
  u_n^{(2)}(b,r) \equiv  u_n(b,b,r)
\end{equation}
or a pair including one point mass
\begin{equation}\label{eq_force_b:1}
  u_n^{(1)}(b,r) \equiv
  -\frac{2}{\pi} \int_0^\infty j_0(kr) \;
  \mathcal{W}_n\left(\frac{kb}{2n}\right) 
  dk\;.
\end{equation}

Note that as $n\rightarrow\infty$ at fixed $b$,
$W_n(b,r)\rightarrow\delta^{(3)}(r)$.  Therefore, even at fixed scale length
$b$, the force profile approaches that of a point mass given this choice of
normalization.

\subsection{W-shape Softening}\label{sec_wshape}

As noted above, the scale length $b$ is not quite the softening
length: the effective softening also depends upon the smoothing index
$n$ when we define the cloud profile to vanish exactly at $r>b/2$.
For convenience, we therefore choose to redefine the softening length
to absorb this $n$ dependence.  We may do so by rescaling the
softening length, writing 
\begin{equation}\label{eq_wshape_scaling}
  b = K_{pn}^{(2)} \eps_W
\end{equation}
where the values of the coefficients $K_{pn}^{(2)}$ for typical values $n=1-4$
are given in the table in Appendix~\ref{app_asm}.  The constant is chosen to
make the interaction potential between two clouds of equal smoothing
scale~$\eps_W$ at zero separation depend only on $\eps_W$ and not on $n$,
i.e.\
\begin{equation}
  u_n^{(2)}(b,r=0)=-\frac{1}{\eps_W}\;.
\end{equation}

\begin{figure}
\includegraphics[scale=0.6]{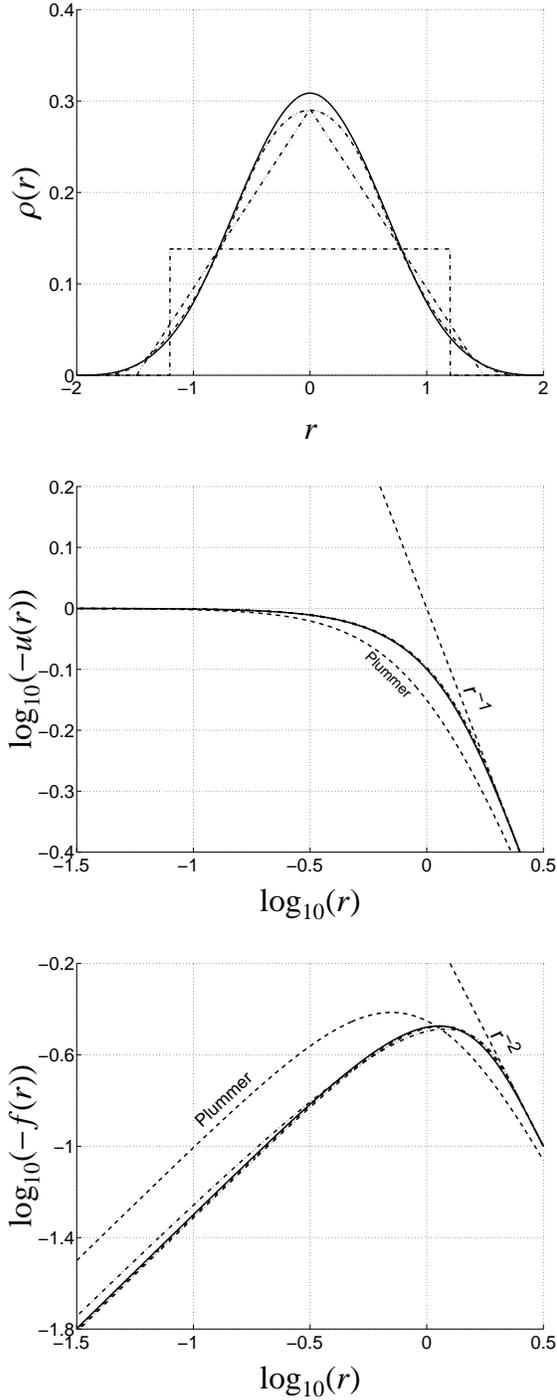}
\caption{\label{fig_pot} Density, force and potential profiles
  for $\eps_W=1$ and $n=1,2,3,4$.  Plummer ($\eps_P=1$)
  profiles are shown. 
 }
\end{figure}

Figure \ref{fig_pot} illustrates the density, potential, and force profiles
for various $n$ at fixed $\eps_W$, for pairs of identical particles.

As may be apparent from the figure, as we take the limit $n\rightarrow\infty$
at fixed $\eps_W$, the density profile converges to a gaussian discussed in
section~\ref{sec_gaussian}. 
Indeed, as suggested by table in Appendix~\ref{app_asm} the coefficient in
Eqn.~(\ref{eq_wshape_scaling}) scales roughly as $K_{pn}^{(2)} \propto
\sqrt{n}$.  Assuming this scaling for~$n\rightarrow\infty$, using
Eqns.(\ref{eq_wshape_scaling}), (\ref{eq_ftwk}), ~(\ref{eq_ftwkfull}), and
taking the limit $n\rightarrow\infty$ we arrive to the gaussian clouds
Eqn.~(\ref{eq_gausswk}).

\begin{figure}
\includegraphics[scale=0.65]{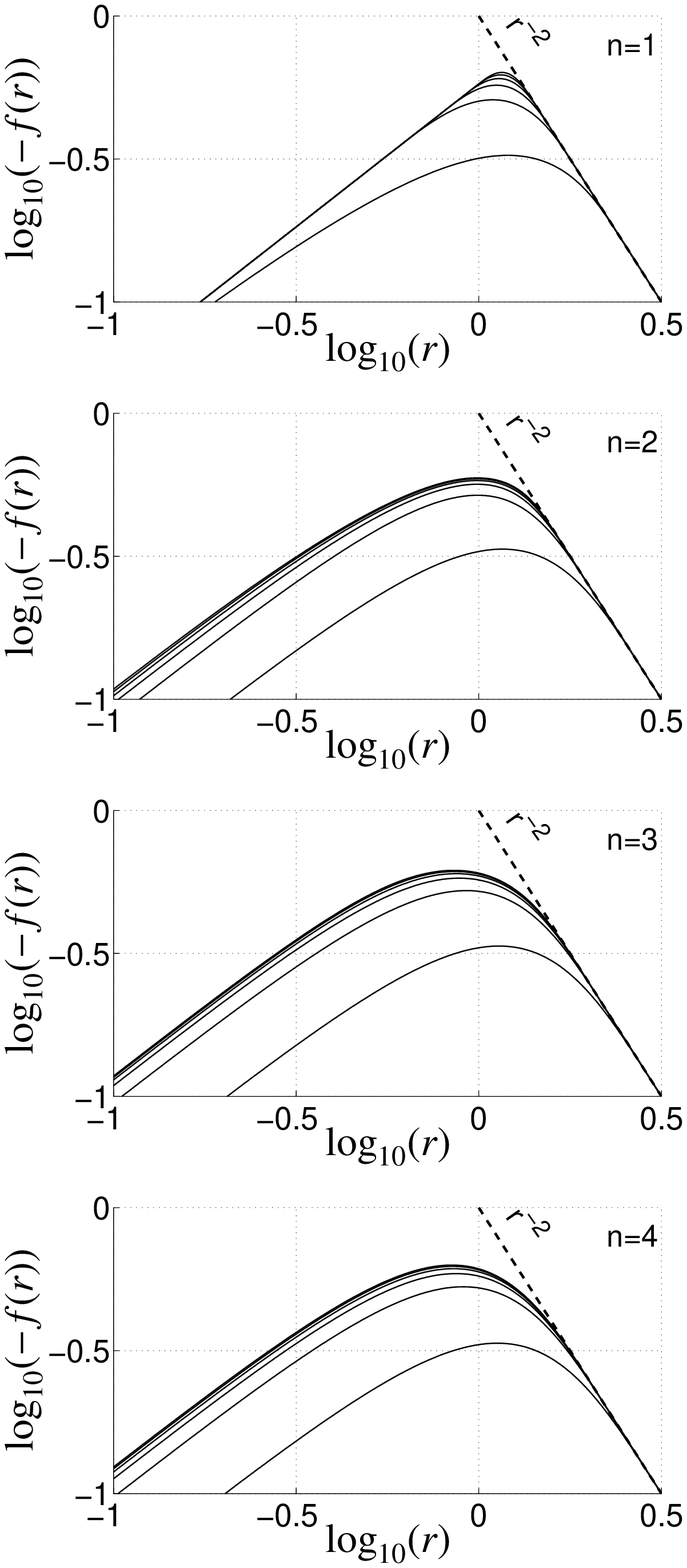}
\caption{\label{fig_forc_gen} Interparticle force laws in pair of particles of
  unequal softening lengths $\eps_W=1$ and $\eps_W=1/q$.  Solid lines
  correspond to different values of~$q=1,\ldots,6$ from bottom to top.  }
\end{figure}

Next we consider the pairwise force between $W_n$ shaped particles of unequal
softening length.  Figure~\ref{fig_forc_gen} shows examples of the force laws
between particles of smoothing scales $\eps_W=1$ and $\eps_W=1/q$; the
different panels show different smoothing indices $n$.  As the ratio
$q\rightarrow\infty$, this approaches the interaction between a $W_n$ cloud
and a point particle.  As is apparent from the figure, the force profiles
quickly converge to the asymptotic ($n=\infty$) behaviors for both~$n>1$
and~$q>1$.


\subsection{Spline Softening}\label{sec_spline}
As noted above, our $n=4$ density shape corresponds to the spline kernels used
in, e.g., SPH \cite{Monaghan85,Springel05,Price07} smoothing.  However the spline
force law does not exactly match ours.  This is because the spline force used
commonly in the literature does not correspond to the force law between two
$W_4$ clouds, but rather to the force law between a $W_4$ cloud and a point
particle.

In spline softening, interparticle potential and force laws follow from
Appendix~\ref{app_onepar}. Zero separation interaction potential becomes
\begin{equation}
  u_4^{(1)}(b,r=0) = -\frac{K_{p4}^{(1)}}{b} = -\frac{28}{5 b} \equiv -\frac{1}{\eps_{\rm spline}}\;,
\end{equation}
where we have used equation~(\ref{eq_asmpt}) and scaling
\begin{equation}
  b=K_{p4}^{(1)} \eps_{\rm spline}\;.
\end{equation}

It may first appear given the discussion in section~\ref{sec_plummer} that
spline softening can be viewed as force between the spherically symmetric
clouds of shapes~$\sqrt{W_4}$. We can show however that spline softening is
inconsistent with such model.  Indeed, equations~(\ref{eq_ftwk})
and~(\ref{eq_ftwkfull}) show $W_4$ is negative for some~$k$, hence
$\sqrt{W_4}$ is imaginary.  On the other hand, equation~(\ref{eq_wreal}) shows
that the fourier component of any spherically symmetric cloud is real. By
contradiction this proves that there is no possible solution for the density
shape~$\rho(r)$ that for a pair of identical particles of this shape would
under gravitational interaction give us the force law used in the spline force
softening.  

Whether or not this conclusion brings important consequences for simulations
that use spline softening is debatable and is open for future tests.  Using
the extended Plummer, W-shape or gaussian softenings however allows one to
immediately resolve this uncertainty. 

\section{Gaussian Softening}\label{sec_gaussian}
Gaussian smoothing
\begin{eqnarray}\label{eq_gausswk}
  W_G(\eps_G, r)&=&\frac{\exp  ( - \pi r^2 / (2\eps_G^2)  )  }{2^{3/2}\eps_G^3} \\
  \mathcal{W}_G(\eps_G, k)&=&\exp\left[-(k\eps_G)^2/2\pi\right].
\end{eqnarray}
allows the simplest expression for softening between clouds of different
smoothing scale and, as shown in section~\ref{sec_wshape}, is the
$n\rightarrow\infty$ limit of softening with our $W_n$ shapes.  For the
interaction potential between two gaussian clouds of softenings $\eps_1$ and
$\eps_2$ whose centers are separated by distance $r$ this gives
\begin{equation}
  U_G(\eps_1,\eps_2,r)=-\frac{m_1 m_2}{r}{\rm erf}\left(\frac{\sqrt{\pi}r}{2\eps_{\rm sym}}\right)\;,
\label{gaussian_pot}
\end{equation}
where
\begin{equation}\label{thumb}
  \eps_{\rm sym} = \eps_{\rm sym}(\eps_1,\eps_2) = \sqrt{\frac{1}{2}\left(\eps_1^2+\eps_2^2\right)}\;.
\end{equation}

We have found that interaction potential in pair of gaussian clouds of
softening scales~$\eps_1$ and~$\eps_2$ equals interaction potential between
identical gaussian clouds each having softening scale~$\eps_{\rm sym}$.

This simple prescription in Eqn.~(\ref{thumb}) however does not generalize to
most other commonly used shapes. For a general cloud shape~$W(k)$ and general
symmetric combination~$\eps_{\rm sym}(\eps_1,\eps_2)$ the prescription is
valid only if
\begin{equation}\label{wewe}
  W(\eps_{\rm sym}(\eps_1,\eps_2) ,k) = [W(\eps_1,k)W(\eps_2,k)]^{1/2}
\end{equation}
for all $k$, as can be seen from Eqn.~(\ref{eqn:poten}). The condition is
exact for gaussian clouds and~$\eps_{\rm sym}$ given by Eqn.~(\ref{thumb}),
but is not satisfied for other commonly used shapes: Plummer, spline or
W-shapes.

\section{Relations between Softenings}\label{sec_relat}

In this section, we turn to the relation between our proposed W-shape (we
assume $n=4$), extended Plummer laws and the more familiar spline, Plummer
and gaussian force softenings.

\begin{figure}
\includegraphics[scale=0.65]{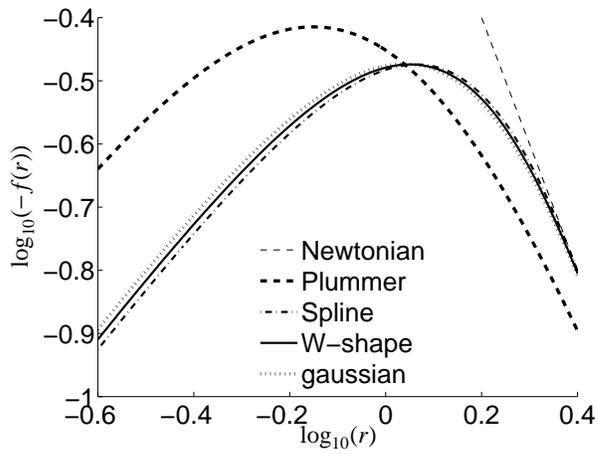}
\caption{\label{fig_forc_compar} 
  Plummer, W-shape, spline and gaussian force laws for $\eps =1$.
}
\end{figure}

How different, in practice, are our proposed  profiles
from previously used laws?  To answer this, we first must normalize the
various profiles to match each other as closely as possible.  We do so by
matching zero separation interaction potentials in pairs of identical particles. Normalized in this way
\begin{equation}
  \eps_G = \eps_P = \eps_W = \eps_{\rm spline} \equiv \eps\;,
\end{equation}
which allows us to drop the subscripts. We see from
figure~\ref{fig_forc_compar} that our $W$-shape profiles are quite close to
spline and gaussian softening for the same~$\eps$ but diverge from Plummer
profile.

\begin{figure}
\includegraphics[scale=0.65]{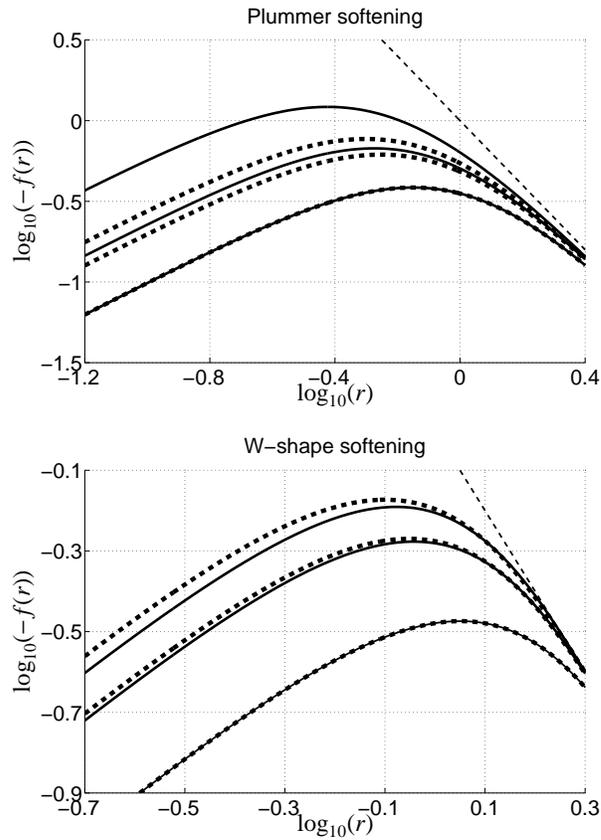}
\caption{\label{fig_compar_adap} Force law in pairs of softenings $\eps_1=1$ and
  $\eps_2 = 1/q$, for $q=1,2,\infty$ in bottom to top order on each plate.  {\it 
    Solid lines}: found by double integration over particle shapes; {\it
    Dashed lines}: found by using the prescription $\eps =
  \sqrt{(\eps_1^2+\eps_2)/2}$. The solid and dashed curves visibly overlap for $q=1$.}
\end{figure}

The close coincidence of W-shape and spline curves is consistent with the idea
that our W-shapes reach gaussian in the limit $n\rightarrow \infty$.  It may
therefore appear first that the prescription $\eps =
\sqrt{(\eps_1^2+\eps_2^2)/2}$ of Eqn.~(\ref{thumb}) is the consistent way to
generalize the interparticle laws to pairs of unequal softenings.  As a
measure of consistency, in figure \ref{fig_compar_adap} we plot the force laws
found using this prescription against the ``correct'' force laws found in the
result of double integration over particle shapes.  The coincidence is
identical for gaussian softening (not shown in figure), as we know from
section~\ref{sec_gaussian}.

For Plummer force law the solid lines are found by numerical integration using
Eqns.~(\ref{eqn:poten}) and~(\ref{plummer:w}) and the dashed lines are found
by plugging $\eps_P = \sqrt{(\eps_1^2+\eps_2^2)/2}$ into Eqn.~(\ref{plummer}).
From the plot we find this prescription leads to up to $52 \%$ relative
systematic inconsistency force.

For W-shape softening this prescription results in at most $10 \%$ systematic
increase over the self-consistent Poisson gravity force law. The latter is
however easily computable using the expressions we provided, hence there is no
advantage in using the simplification in the first place.

\section{Numerical Implementation}\label{sec_numeric}

The documented {\tt WSHAPE} package (this paper uses version~$1.0$) is
available at {\tt http://www.gracos.org/wshape}.  We provide numerical
C-implementation of the potential and force laws for W-shape and extended
Plummer. Also provided within the package is the procedure used to arrive to
expressions in Appendix~\ref{app_tables} and values of coefficients in
Table~\ref{tab_Ks}.

\section{Conclusions}\label{sec:concl}

An N-body simulation is a simulation of massive particles under the influence
of gravity.  In cosmological simulation interparticle force is softened to
avoid hard collisions. Simulation particles have often been viewed as
particles of fixed density shapes in previous literature (e.g. expression for
$\rho(r)$ in Eqn.~(5) of \cite{Springel05}).  To ensure that the nature of
interaction between particles in simulations remains gravitational we adpot
this interpretation in which force softening corresponds to interaction
between cloudlike particles of a fixed density shape.

We prove that the Plummer interparticle force softening is consistent with
this force softening model and generalize the Plummer force law into the case
of unequal softening scales $\eps_1$ and~$\eps_2$.  

Interestingly, we mathematically prove that the previosuly used in literature
(e.g.  \cite{Springel05}) method of {\it spline softening} is inconsistent
with gravitational interaction in the sense that there is no possible solution
for the density shape~$\rho(r)$ that for a pair of identical particles of this
shape would under gravitational interaction give us the force law used for
interparticle force law in spline softening.

We provide the closed form solution for potential and force laws between
widely known $W_n$-cloud shapes of general softenings~$\eps_1$ and $\eps_2$.

Our generalized interparticle force laws should be useful for N-body
simulations with adaptive mass refinement, in which particles of different
mass interact gravitationally.  Examples include simulations of the first
collapsed objects in the universe \cite{Diemand05}, the ``Via Lactea''
simulation \cite{Diemand07}, pure gravity extension of particle
splitting~\cite{mes05}, and simulations of dark matter that includes pointlike
objects, e.g.  stars or supermassive black holes.

As an easy way to extend the force law~$f(\eps,r)$ into pairs of particles of
unequal softenings~$\eps_1$ and $\eps_2$ one would plug their symmetric
combination of ~$\eps_1$ and $\eps_2$ as the softening scale~$\eps$. We found
that using the simple gaussian motivated recipe~$\eps =
\sqrt{(\eps_1^2+\eps_2^2)/2}$ for simulations with adaptive mass resolution
generally leads to significant systematic uncertainties for the pairs of
particles of unequal softenings. The overall effect of these errors can be
established by numerical convergence tests similar to \cite{Heitmann05}.
However the problem is immediately resolved by using our proposed profiles.

For convenience, we also made a numeric implementation of our analytic
expressions.  The most efficient way to use these laws in an N-body simulation
would be to pre-compute the interaction law once before running the
simulation, and then to interpolate by look-up from the precomputed table.
The {\tt GRACOS} package \cite{sb05} will implement these profiles for
adaptive softening.

\section*{Acknowledgments}
I thank Neal Dalal for suggesting this topic, and for
helpful discussions, in particular for the derivation of the Gaussian limit of
the $W_n$ shapes. I thank Pascal Vaudrevange, Pat McDonald, John Dubinski,
Norm Murray, Ue-Li Pen, Latham Boyle, Sergei Shandarin and Lev Kofman for helpful discussions. I thank E.
Bertschinger for his teaching and guidance on N-body codes.  This work was
supported by the Canadian Institute for Theoretical Astrophysics (CITA) and
the Natural Sciences and Engineering Research Council of Canada (NSERC).

\newcommand\apjs{\rm {ApJS}}%

\newcommand\apj{\rm {ApJ}}%

\newcommand\nat{\rm {Nature}}%
\newcommand\aap{\rm {A\&A}}%
\newcommand\mnras{\rm {MNRAS}}%


\appendix

\section{Force and Potential Law in a Pair of $W_n$ Cloud Shapes}\label{app_force}

This section provides closed form expressions (for $n=1,2,3,4$) for
interaction potential laws in a pair of two $W_n$ cloud shapes of scales~$b_1$
and~$b_2$, a pair of two identical $W_n$-shape particles of scale~$b$, and a
pair consisting of an $W_n$-shape particles and a point-like
particle.  Interparticle force laws follow immediately by differentiation with
respect to separation~$r$ and flipping the sign.

\subsection{Analytic Expression}\label{app_analyt}

\subsubsection {Two Parametric Form}\label{app_twopar}
Potential law~$u_n(b_1,b_2,r)$ for interaction in a pair of two $W_n$-shape
particles of scales~$b_1$ and~$b_2$ and a fixed~$n$ is given by Eqn.~(\ref{eq_force_b1b2}).
Its closed form solution is given as a finite sum over all integers~$i$
and~$j$, with
\begin{equation}\label{eq_poten_gen}
u_n(b_1,b_2,r) = 
-\frac{1}{A_n r}
\sum_{ij}
A_{nij}(r)
\;
(r/a_1)^{n+2-i}
(r/a_2)^{n+2-j}
\end{equation}
where $a_s \equiv b_s/(2M_n)$ for $s=1,2$, and the values of positive integers
$M_n$ and $A_n$ are given for each~$n$ in Appendix~\ref{app_tables}.

Coefficients $A_{nij}(r)$ depend on~$r$ through the Heaviside function~$H_0(r)$ via
summation over all integers~$p$ and~$q$
\begin{equation}\label{eq_Anij}
  A_{nij}(r) = \sum_{pq} \frac{1}{2} \left[ C_{npq}(i,j)+C_{npq}(j,i) \right]\, H(p,q,r)\;,
\end{equation}
where $H(p,q,r) \equiv H_0(r-(p a_1 + q a_2))$ for integers~$p$ and~$q$, and
$C_{npq}(i,j)$ are constant non-zero integer coefficients, whose complete set
for each~$n$ is presented in Appendix~\ref{app_tables}. The total number of
terms in these expressions is finite.  The potential is symmetric with
respect to~$b_1$ and~$b_2$, as should be.

\subsubsection {One Parametric Forms}\label{app_onepar}
Potential laws for interaction in a pair of two identical $W_n$-shape
particles of scale~$b$ (case $c=2$), or one $W_n$-shape particle of scale $b$
and a point-like particle (case $c=1$) are found in Eqn.~(\ref{eq_force_b:1})
and~(\ref{eq_force_b:2}). Their closed form solutions are given by
\begin{equation}\label{eq_poten_iden}
u_n^{(c)}(b,r) = 
-\frac{1}{A_n^{(c)} r}
\sum_{i}
A_{ni}^{(c)}(r)\;(r/a)^{(n+c-1)c+2-i}\;,
\end{equation}
where $a \equiv b/(2M_n)$. The values of positive integers $M_n$ and
$A_n^{(c)}$ are given for each~$n$ and~$c=1,2$ in Appendix~\ref{app_tables}.
Coefficients $A_{ni}^{(c)}(r)$ depend on~$r$ through the Heaviside
function~$H_0(r)$ via
\begin{equation}\label{eq_Anijc}
  A_{ni}^{(c)}(r) = \sum_{p} C_{np}^{(c)}(i)\, H(p,r)
\end{equation}
where $H(p,r) \equiv H_0(r-p a)$ for an integer~$p$, and $C_{np}^{(c)}(i)$ are
constant non-zero integer coefficients, whose complete set for each~$n$ is
also presented in Appendix~\ref{app_tables}.

As an exercise compute~$u_4^{(1)}(4h,r)$, using the above expressions and
table values. You should arrive to the right hand side of Eqn.(A2) of
\cite{Price07}, which is the potential kernel in the traditional spline
softening described in Section~\ref{sec_spline}.

\subsection{Asymptotic Expressions}\label{app_asm}

\begin{table}
\caption{\label{tab_Ks} Table of coefficients for Eqn.(\ref{eq_asmpt})}
\begin{minipage}{10cm}
\begin{displaymath}
\begin{array}{@{\extracolsep{9mm}}c@{\hspace{4mm}}|@{\hspace{2mm}}cccc}
\multicolumn{5}{l}{\mbox{Pair: one W-shape and one point particle ($c=1$)}}\\
n                   & 1     &    2    & 3       &    4  \\
\hline
\rule[-1mm]{0mm}{5mm}
K_{pn}^{(1)}       &  3    &  4      &  39/8   & 28/5  \\
\rule[-1mm]{0mm}{5mm}
K_{fn}^{(1)}       &  8    & 32      &  54     & 256/3 \\[5pt]
\hline
\end{array}
\end{displaymath}
\noindent
\begin{displaymath}
\begin{array}{@{\extracolsep{9mm}}c@{\hspace{4mm}}|@{\hspace{2mm}}cccc}
\multicolumn{5}{l}{\mbox{Pair of identical W-shape particles ($c=2$)}}\\
  n                   & 1     &    2    & 3       &    4         \\
\hline
\rule[-1mm]{0mm}{5mm}
  K_{pn}^{(2)}       &  12/5 &  104/35 &  124/35 & 70016/17325  \\
\rule[-1mm]{0mm}{5mm}
  K_{fn}^{(2)}       &  8    & 64/5    & 774/35  & 31424/945 \\[5pt]
\hline
\end{array}
\end{displaymath}
\end{minipage}
\end{table}

Potential laws in section \ref{app_analyt} are pure Newtonian beyond
separation distance when particle last overlap.  Indeed~$u_n(b_1,b_2,r) =
-1/r$ at $r \ge (b_1+b_2)/2$, and~$u_n^{(c)}(b,r) = -1/r$ at $r \ge c b/2$.
On the other hand, for small separations we have analytically for $c=1,2$
\begin{equation}\label{eq_asmpt}
  \begin{array}{cc}
    \begin{array}{c}
      u_n^{(c)}(b,r) = -K_{pn}^{(c)}/b\\
      f_n^{(c)}(b,r) = -K_{fn}^{(c)}r/b^3
    \end{array} 
    ,\qquad\mbox{as}
    &
    r \ll b\;,
  \end{array}
\end{equation}
where numerical values of coefficients are given in Tables~\ref{tab_Ks}.

\begin{figure}
  \includegraphics[scale=0.6]{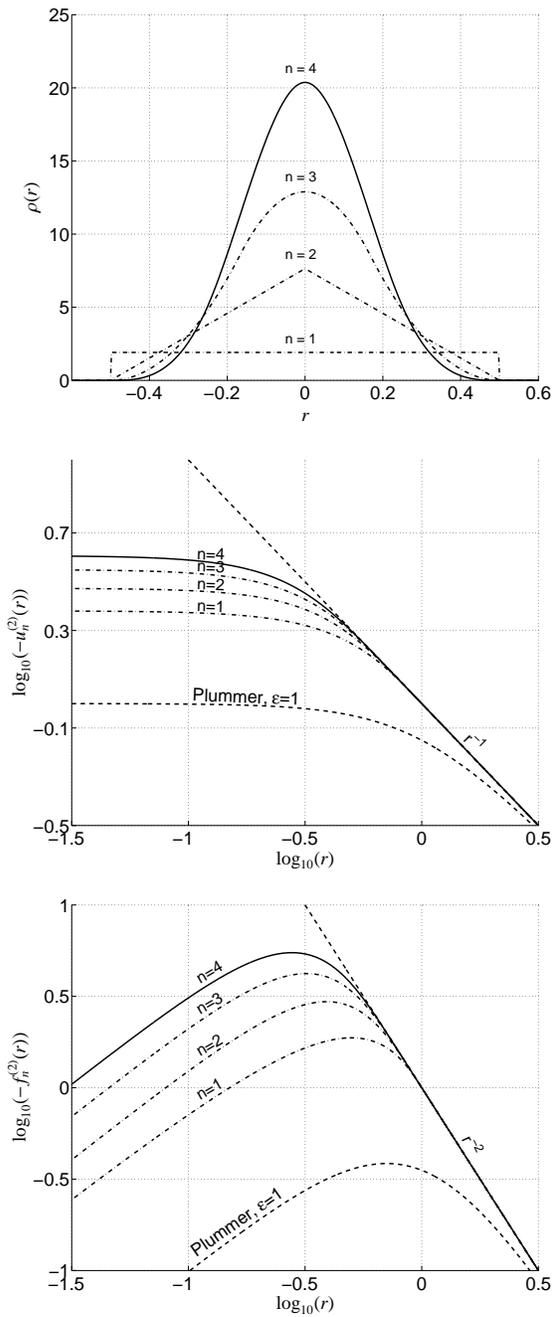}
  \caption{\label{fig_noscale} 
    Density, force and potential laws for~$b=1$ for unscaled $W_n$-shapes for $n=1,2,3,4$.
    Plummer law for~$\eps=1$ is also shown.
  }
\end{figure}

Figure~\ref{fig_noscale} shows the density, potential and force laws for $n=1-4$
for pair of identical particles of smoothing scale~$b=1$. Note the different
amplitudes for the potential and linear force law at small $r$ for different
$n$, which is consistent with growing values of $K_{pn}^{(2)}$ and
$K_{fn}^{(2)}$ for $n=1,2,3,4$.

\section{Tables of Coefficients}\label{app_tables}
This subsection provides tables of coefficients for analytic expressions for
potentials in Section~\ref{app_analyt} for~$n = 1,2,3,4$.  The procedure used
to find these table values is given in Appendix~\ref{sec_numeric}.

\subsection{Spherical Top-hat -- Shaped Particles ($W_1$-Shape)}\label{app_s1}

Values $n=1$, $M_n=1$ are applicable for the entire subsection~\ref{app_s1}.

\subsubsection{Point-like Test Particle}

\begin{displaymath}
A_n^{(1)}=2
\end{displaymath}

\noindent
The following table lists values of coefficients~$C_{np}^{(1)}(i)$;
the numbers, labeling the rows and columns denote~$p$
and~$i$.
\begin{displaymath}
\begin{array}{@{\extracolsep{1mm}}c@{\hspace{4mm}}|@{\hspace{2mm}}ccc}
\multicolumn{4}{c}{C_{np}^{(1)}(i)}\\
        & 0  & 2  & 3 \\\hline
0       & -1 & 3  &   \\
1       & 1  & -3 & 2 \\
\end{array}
\end{displaymath}

\subsubsection{Identical Particles}

\begin{displaymath}
A_n^{(2)}=160
\end{displaymath}

\noindent
The following table lists values of coefficients~$C_{np}^{(2)}(i)$;
the numbers, labeling the rows and columns denote~$p$
and~$i$.
\begin{displaymath}
\begin{array}{@{\extracolsep{1mm}}c@{\hspace{4mm}}|@{\hspace{2mm}}ccccc}
\multicolumn{6}{c}{C_{np}^{(2)}(i)}\\
        & 0  & 2   & 3   & 5    & 6   \\\hline
0       & -1 & 30  & -80 & 192  &     \\
2       & 1  & -30 & 80  & -192 & 160 \\
\end{array}
\end{displaymath}

\subsubsection{General Case}

\begin{displaymath}
A_n=160
\end{displaymath}

\noindent
The following two tables list values of coefficients~$C_{n00}(i,j)$ and
$C_{n11}(i,j)$; 
the numbers, labeling the rows and columns denote~$i$ and~$j$.
\begin{displaymath}
\begin{array}{c}
\begin{array}{c@{\hspace{4mm}}|@{\hspace{2mm}}cccccc}
\multicolumn{7}{c}{C_{n00}(i,j)}\\
        & 0 & 2   & 3   & 4   & 5   & 6   \\\hline
0       & 1 & -30 & -80 & -90 & -48 & -10 \\
2       &   & 90  & 240 & 90  &     &     \\
3       &   &     & -80 &     &     &     \\
\end{array}
\\
\begin{array}{c@{\hspace{4mm}}|@{\hspace{2mm}}cccccc}
\multicolumn{7}{c}{C_{n11}(i,j)}\\
        & 0 & 2   & 3    & 4   & 5  & 6   \\\hline
0       & 1 & -30 & 80   & -90 & 48 & -10 \\
2       &   & 90  & -240 & 90  &    &     \\
3       &   &     & 80   &     &    &     \\
\end{array}
\end{array}
\end{displaymath}

\noindent
The rest of the non zero coefficients in Eqn.~(\ref{eq_Anij}) are given by the
following relations
\begin{displaymath}
\begin{array}{l}
C_{n,-1,1}(i,j) = (-1)^{1+i}\; C_{n11}(i,j)\\
C_{n,1,-1}(i,j) = (-1)^{1+j}\; C_{n11}(i,j)\;.
\end{array}
\end{displaymath}

\subsection{Cone -- Shaped Particles ($W_2$-Shape)}\label{app_s2}

Values $n=2$, $M_n=1$ are applicable for the entire subsection~\ref{app_s2}.

\subsubsection{Point-like Test Particle}

\begin{displaymath}
A_n^{(1)}=1
\end{displaymath}

\noindent
The following table lists values of coefficients~$C_{np}^{(1)}(i)$; 
the numbers labeling the rows and columns denote~$p$ and~$i$.
\begin{displaymath}
\begin{array}{c@{\hspace{4mm}}|@{\hspace{2mm}}cccc}
\multicolumn{4}{c}{C_{np}^{(1)}(i)}\\
        & 0  & 1 & 3  & 4 \\\hline
0       & 1 & -2 & 2 \\
1       & -1 & 2 & -2 & 1 \\
\end{array}
\end{displaymath}

\subsubsection{Identical Particles}

\begin{displaymath}
A_n^{(2)}=140
\end{displaymath}

\noindent
The following table lists values of coefficients~$C_{np}^{(2)}(i)$;
the numbers, labeling the rows and columns in the following table, denote~$p$
and~$i$ respectively.
{
  \small
\begin{displaymath}
\begin{array}{c@{\hspace{4mm}}|@{\hspace{2mm}}
    c@{\extracolsep{6pt}}
    c@{\extracolsep{6pt}}
    c@{\extracolsep{6pt}}
    c@{\extracolsep{6pt}}
    c@{\extracolsep{6pt}}
    c@{\extracolsep{6pt}}
    c@{\extracolsep{6pt}}
    c@{\extracolsep{6pt}}
    c@{\extracolsep{6pt}}
  } 
        \multicolumn{10}{c}{C_{np}^{(2)}(i)}\\
        & 0  & 1  & 2   & 3    & 4    & 5    & 6    & 7    & 8   \\\hline
0       & 3  & -8 & -14 & 56   &      & -112 &      & 208 \\
1       & -4 & 16 &     & -112 & 280  & -336 & 224  & -80 & 12 \\
2       & 1  & -8 & 14  & 56   & -280 & 448  & -224 & -128 & 128 
\end{array}
\end{displaymath}
}

\subsubsection{General Case}

\begin{displaymath}
A_n=140
\end{displaymath}          

\noindent
The following three tables list values of coefficients~$C_{n00}(i,j)$,
$C_{n01}(i,j)$, and~$C_{n11}(i,j)$; 
the numbers, labeling the rows and columns
denote~$i$ and~$j$ respectively.  
{ \small
  \begin{displaymath}
\begin{array}{c@{\hspace{4mm}}|@{\hspace{2mm}}ccccccccc}
\multicolumn{9}{c}{C_{n00}(i,j)}\\
        & 0 & 1  & 3    & 4    & 5    & 6    & 7   & 8 \\\hline
0       & 1 & -8 & 56   & 140  & 168  & 112  & 40  & 6 \\
1       &   & 14 & -140 & -280 & -252 & -112 & -20 &   \\
3       &   &    & 140  & 280  & 84   &      &     &   \\
4       &   &    &      & -70  &      &      &     &   \\
\\[3pt]
\multicolumn{9}{c}{C_{n01}(i,j)}\\
        & 0  & 1 & 3   & 4   & 5    & 6   & 7   & 8 \\\hline
0       & -2 & 8 & -56 & 140 & -168 & 112 & -40 & 6 \\
\end{array}
\end{displaymath}          
\begin{displaymath}
\begin{array}{c@{\hspace{4mm}}|@{\hspace{2mm}}ccccccccc}
\multicolumn{9}{c}{C_{n11}(i,j)}\\
        & 0 & 1  & 3    & 4    & 5    & 6    & 7   & 8  \\\hline
0       & 1 & -8 & 56   & -140 & 168  & -112 & 40  & -6 \\
1       &   & 14 & -140 & 280  & -252 & 112  & -20 &    \\
3       &   &    & 140  & -280 & 84   &      &     &    \\
4       &   &    &      & 70   &      &      &     &    \\
\end{array}
\end{displaymath}          
}

\noindent
The rest of the non zero coefficients are given by the
following relations.
\begin{displaymath}
\begin{array}{llll}
C_{n10} (j,i)  & = &        & C_{n01}(i,j)\\[1pt]
C_{n,1,-1}(i,j) & = & (-1)^j & C_{n11}(i,j) \\
C_{n,-1,1}(i,j) & = & (-1)^i & C_{n11}(i,j) 
\end{array}
\end{displaymath}

\subsection{TSC-Shape Particles ($W_3$-Shape)}\label{app_s3}

Values $n=3$, $M_n=3$ are applicable for the entire subsection~\ref{app_s3}.

\subsubsection{Point-like Test Particle}

\begin{displaymath}
A_n^{(1)}=160
\end{displaymath}

\noindent
The following table lists values of coefficients~$C_{np}^{(1)}(i)$; 
the numbers, labeling the rows and columns denote~$p$ and~$i$ respectively.
\begin{displaymath}
\begin{array}{c@{\hspace{4mm}}|@{\hspace{2mm}}ccccc}
\multicolumn{6}{c}{C_{np}^{(1)}(i)}\\
        & 0  & 1   & 2   & 4    & 5   \\\hline
0       & 2  &     & -20 & 130        \\
1       & -3 & 10  & -10 & 5    & -2  \\
3       & 1  & -10 & 30  & -135 & 162 
\end{array}
\end{displaymath}

\subsubsection{Identical Particles}

\begin{displaymath}
A_n^{(2)}=12902400
\end{displaymath}

\noindent
Table~\ref{tab_iden3} lists values of coefficients~$C_{np}^{(2)}(i)$.

\subsubsection{General Case}

\begin{displaymath}
A_n=12902400
\end{displaymath}          

\noindent
Table~\ref{tab_gen3} lists values of coefficients~$C_{n11}(i,j)$.
The rest of the non zero coefficients~$C_{npq}(i,j)$ are given by the
following relations
\begin{displaymath}
\hspace{-1cm}
\begin{array}{c|c}
\begin{array}{lccl}
C_{n,1,-1}(i,j) &  = &   (-1)^{1+j}               & C_{n11}(i,j), \\
C_{n,-1,1}(i,j) &  = &   (-1)^{1+i}               & C_{n11}(i,j), \\
C_{n,1,-3}(i,j) &  = &   (-1)^j     M_n^{-1+j}    & C_{n11}(i,j), \\
C_{n,-3,1}(i,j) &  = &   (-1)^i     M_n^{-1+i}    & C_{n11}(i,j), \\
C_{n,3,-1}(i,j) &  = &   (-1)^j     M_n^{-1+i}    & C_{n11}(i,j), \\
C_{n,-1,3}(i,j) &  = &   (-1)^i     M_n^{-1+j}    & C_{n11}(i,j), \\
\end{array}
\quad
&
\quad
\begin{array}{lccl}
C_{n13}(i,j)    & =  & - M_n^{-1+j}              & C_{n11}(i,j), \\
C_{n31}(i,j)    & =  & - M_n^{-1+i}              & C_{n11}(i,j), \\
C_{n33}(i,j)    & =  &   M_n^{-2+i+j}            & C_{n11}(i,j), \\
C_{n,3,-3}(i,j) & =  &   (-1)^{1+j} M_n^{-2+i+j}  & C_{n11}(i,j), \\
C_{n,-3,3}(i,j) & =  &   (-1)^{1+i} M_n^{-2+i+j}  & C_{n11}(i,j). 
\end{array}
\end{array}
\end{displaymath}

\subsection{Cubic Spline - Shaped Particles ($W_4$\,-Shape)}\label{app_s4}

Values $n=4$, $M_n=2$ are applicable for the entire subsection~\ref{app_s4}.

\subsubsection{Point-like Test Particle}

\begin{displaymath}
A_n^{(1)}=30
\end{displaymath}

\noindent
The following table lists values of coefficients~$C_{np}^{(1)}(i)$; the
numbers, labeling the rows and columns denote~$p$ and~$i$.
\begin{displaymath}
\begin{array}{c@{\hspace{4mm}}|@{\hspace{2mm}}cccccc}
\multicolumn{7}{c}{C_{np}^{(1)}(i)}\\
        & 0  & 1   & 2   & 3   & 5   & 6\\\hline
0       & -3 & 9   &     & -20 & 42 \\
1       & 4  & -18 & 30  & -20 & 6   & -2 \\
2       & -1 & 9   & -30 & 40  & -48 & 32 \\
\end{array}
\end{displaymath}

\subsubsection{Identical Particles}\label{app_n4iden}

\begin{displaymath}
A_n^{(2)}=1663200
\end{displaymath}

\noindent
Table~\ref{tab_iden4} lists values of coefficients~$C_{np}^{(2)}(i)$.

\subsubsection{General Case}

\begin{displaymath}
  A_n=1663200
\end{displaymath}

\noindent
Table~\ref{tab_gen4} lists values of coefficients~$C_{n00}(i,j)$,
$C_{n01}(i,j)$ and $C_{n11}(i,j)$.  The rest of the
coefficients~$C_{npq}(i,j)$ are given by the following relations:
\begin{displaymath}
\hspace{-1cm}
\begin{array}{c|c}
\begin{array}{lccl}
C_{n10}(j,i)    & =  &                            &  C_{n01}(i,j)  \\      
C_{n02}(i,j)    & =  &  - M_n ^  {-2+i+j}          &  C_{n01}(i,j)  \\
C_{n20}(j,i)    & =  &  - M_n ^  {-2+i+j}          &  C_{n01}(i,j)  \\
C_{n12}(i,j)    & =  &  - M_n ^  {-2+j}            &  C_{n11}(i,j)  \\
C_{n21}(i,j)    & =  &  - M_n ^  {-2+i}            &  C_{n11}(i,j)  \\
C_{n22}(i,j)    & =  &    M_n ^  {-4+i+j}          &  C_{n11}(i,j)  \\      
C_{n,1,-1}(i,j) & =  &    (-1)^j                  &  C_{n11}(i,j)  \\
\end{array}
\quad
&
\quad
\begin{array}{lccl}
C_{n,-1,1}(i,j) & =  &                      (-1)^i& C_{n11}(i,j)  \\
C_{n,2,-2}(i,j) & =  &   M_n ^  {-4 +i +j}  (-1)^j  & C_{n11}(i,j)  \\
C_{n,-2,2}(i,j) & =  &   M_n ^  {-4 +i +j}  (-1)^i  & C_{n11}(i,j)  \\   
C_{n,1,-2}(i,j) & =  & - M_n ^  {-2 +   j}  (-1)^j  & C_{n11}(i,j)  \\
C_{n,-2,1}(i,j) & =  & - M_n ^  {-2 +i   }  (-1)^i  & C_{n11}(i,j)  \\   
C_{n,2,-1}(i,j) & =  & - M_n ^  {-2 +i   }  (-1)^j  & C_{n11}(i,j)  \\
C_{n,-1,2}(i,j) & =  & - M_n ^  {-2 +   j}  (-1)^i  & C_{n11}(i,j)  \\   
\end{array}
\end{array}
\end{displaymath}

\begin{table*}
\hspace{-1cm}
\begin{minipage}{+15cm}

\caption{\label{tab_iden3} Table of coefficients~$C_{np}^{(2)}(i)$, where $n=3$. Numbers labeling rows and columns in the following table denote~$p$ and~$i$.}
\begin{displaymath}
\hspace{-1cm}
\begin{array}{c@{\hspace{4mm}}|@{\hspace{1mm}}ccccccccccc}
        & 0   & 1    & 2     & 3     & 4       & 5       & 6       & 7        & 8       & 9        & 10 \\\hline
0       & -10 & 80   & 180   & -2880 &         & 48384   &         & -660480  &         & 7618560 \\
2       & 15  & -200 & 810   & 1440  & -26880  & 120960  & -302400 & 468480   & -449280 & 245760   & -58880 \\
4       & -6  & 160  & -1620 & 5760  & 26880   & -387072 & 1935360 & -5406720 & 8847360 & -7864320 & 2883584 \\
6       & 1   & -40  & 630   & -4320 &         & 217728 & -1632960 & 5598720 & -8398080 &          & 10077696 \\
\end{array}
\end{displaymath}

\caption{\label{tab_gen3} Table of coefficients~$C_{n00}(i,j)$ and $C_{n11}(i,j)$, where $n=3$. Numbers, labeling the rows and columns in the following two tables, denote~$i$ and~$j$.}

\begin{displaymath}
\begin{array}{c@{\hspace{4mm}}|@{\hspace{4mm}}ccccccccc}
\multicolumn{10}{c}{C_{n00}(i,j)}\\
        & 0 & 2    & 4       & 5        & 6        & 7        & 8        & 9       & 10     \\\hline
0       & 4 & -360 & 21840   & 161280   & 609840   & 1397760  & 1967400  & 1574400 & 551096 \\
2       &   & 5040 & -327600 & -1612800 & -3659040 & -4193280 & -1967400 &         &        \\
4       &   &      & 2129400 & 10483200 & 7927920  &          &          &         &        \\
5       &   &      &         & -6451200 &          &          &          &         &        \\
\end{array}
\end{displaymath}          
\begin{displaymath}
\begin{array}{c@{\hspace{4mm}}|@{\hspace{4mm}}cccccccccc}
\multicolumn{11}{c}{C_{n11}(i,j)}\\
        & 0 & 1    & 2     & 4     & 5      & 6     & 7     & 8     & 9    & 10  \\\hline
0       & 9 & -120 & 270   & -1260 & 3024   & -3780 & 2880  & -1350 & 360  & -42 \\
1       &   & 360  & -1440 & 5040  & -10080 & 10080 & -5760 & 1800  & -240 &     \\
2       &   &      & 1260  & -6300 & 10080  & -7560 & 2880  & -450  &      &     \\
4       &   &      &       & 3150  & -5040  & 1260  &       &       &      &     \\
5       &   &      &       &       & 1008   &       &       &       &      &     \\
\end{array}
\end{displaymath}          
\end{minipage}
\end{table*}

\begin{table*}
\hspace{-1cm}
\begin{minipage}{+15cm}

\caption{\label{tab_iden4}
Table of coefficients~$C_{np}^{(2)}(i)$, where $n=4$.
The numbers, labeling the rows and columns in the following table, denote~$p$
and~$i$.}
\begin{displaymath}
{
\small
\hspace{-1cm}
\begin{array}{@{\extracolsep{-6pt}}
    c@{\hspace{2mm}}|
    @{\hspace{4mm}}
    c
    c
    c
    c
    c
    c
    c
    c
    c
    c
    c
    c
    c
    c
}
        & 0    & 1      & 2     & 3      & 4     & 5      & 6       & 7        & 8        & 9        & 10       & 11       & 12     \\\hline
0       & 35   & -180   & -165  & 2200   &       & -15840 &         & 95040    &          & -432080  &          & 1680384 \\
1       & -56\;&  504   & -1716 & 1760   & 5940  & -26928 & 53592   & -66528   & 55440    & -31240   & 11484    & -2496    & 244 \\
2       & 28   & -504\; & 3828  & -14960 & 23760 & 50688  & -384384 & 1064448  & -1774080 & 1914880  & -1317888 & 528384   & -94208 \\
3       & -8   & 216    & -2508 & 15840  & -53460& 42768  & 449064  & -2309472 & 5773680  & -8660520 & 7794468  & -3779136\; & 708588 \\
4       & 1    &  -36   & 561   & -4840  & 23760 & -50688\; & -118272 & 1216512  & -4055040 & 7208960  & -6488064 & 1572864  & 1048576 \\
\end{array}
}
\end{displaymath}

\caption{\label{tab_gen4}
Table of coefficients~$C_{n00}(i,j)$, $C_{n01}(i,j)$ and $C_{n11}(i,j)$, where $n=4$.
The numbers, labeling the rows and columns in the following three tables, denote~$i$
and~$j$.
}
{
\begin{displaymath}
\begin{array}{c@{\hspace{4mm}}|@{\hspace{2mm}}ccccccccccccc}
\multicolumn{12}{c}{C_{n00}(i,j)}\\
        & 0 & 1    & 3     & 5       & 6        & 7        & 8       & 9       & 10      & 11      & 12     \\\hline
0       & 9 & -108 & 1320  & -33264  & -166320  & -441936  & -748440 & -838200 & -605880 & -257544 & -49104 \\
1       &   & 297  & -5940 & 116424  & 498960   & 1104840  & 1496880 & 1257300 & 605880  & 128772  &        \\
3       &   &      & 18480 & -388080 & -1108800 & -1473120 & -997920 & -279400 &         &         &        \\
5       &   &      &       & 814968  & 2328480  & 1031184  &         &         &         &         &        \\
6       &   &      &       &         & -831600  &          &         &         &         &         &        \\
\end{array}
\end{displaymath}
}

{
\small
\begin{displaymath}
\begin{array}{c}
\begin{array}{@{\extracolsep{2.8mm}}c@{\hspace{4mm}}|@{\hspace{2mm}}ccccccccccccc}
\multicolumn{13}{c}{C_{n01}(i,j)}\\
        & 0   & 1   & 2    & 3    & 5     & 6     & 7      & 8     & 9     & 10   & 11   & 12 \\\hline
0       & -24 & 216 & -792 & 1320 & -4752 & 11088 & -14256 & 11880 & -6600 & 2376 & -504 & 48 \\
\end{array}
\end{array}
\end{displaymath}
}
{
\small
\begin{displaymath}
\begin{array}{c}
\begin{array}{c@{\hspace{4mm}}|@{\hspace{2mm}}cccccccccccc}
\multicolumn{13}{c}{C_{n11}(i,j)}\\
        & 0  & 1    & 2     & 3      & 5      & 6       & 7      & 8      & 9      & 10    & 11   & 12  \\\hline
0       & 16 & -288 & 1056  & -1760  & 6336   & -14784  & 19008  & -15840 & 8800   & -3168 & 672  & -64 \\
1       &    & 1188 & -7920 & 11880  & -33264 & 66528   & -71280 & 47520  & -19800 & 4752  & -504 &     \\
2       &    &      & 11880 & -31680 & 66528  & -110880 & 95040  & -47520 & 13200  & -1584 &      &     \\
3       &    &      &       & 18480  & -55440 & 73920   & -47520 & 15840  & -2200  &       &      &     \\
5       &    &      &       &        & 16632  & -22176  & 4752   &        &        &       &      &     \\
6       &    &      &       &        &        & 3696    &        &        &        &       &      &     \\
\end{array}
\end{array}
\end{displaymath}
}
\end{minipage}
\end{table*}

\end{document}